\documentclass[aps,showpacs, nofootinbib,floatfix]{revtex4}
\usepackage{graphicx, graphics, color}

\input amssym.tex
\input amssym.def

\newcommand{\be}{\begin{equation}}
\newcommand{\ee}{\end{equation}}
\newcommand{\ben}{\begin{eqnarray}}
\newcommand{\een}{\end{eqnarray}}

\newcommand{\p}{\partial}

\begin{document}

\title{Magnetic field in holographic superconductor with dark matter sector}
%%%%%%%%%%%%%%%%%%%%%%%%%%%%%%%%%%%%%%%%%%%%%%%%%%%%%%%%%%%%%%
%\author{Gary W.Gibbons}
%\affiliation{DAMTP, Centre for Mathematical Sciences,  \protect \\
%University of Cambridge\protect \\
%Wilberforce Road, Cambridge, CB3 0WA, UK\protect \\
%g.w.gibbons@damtp.cam.ac.uk} 

\author{{\L}ukasz Nakonieczny}
\email{Lukasz.Nakonieczny@fuw.edu.pl}
\email{Lukasz.Nakonieczny@fuw.edu.pl}
\affiliation{Institute of Theoretical Physics, Faculty of Physics, University of Warsaw \protect \\
ul.~Pasteura 5,~02-093 Warszawa, Poland }

\author{Marek Rogatko} 
\email{rogat@kft.umcs.lublin.pl,
marek.rogatko@poczta.umcs.lublin.pl }
\author{Karol I. Wysokinski}
\email{karol.wysokinski@umcs.pl}
\affiliation{Institute of Physics \protect \\
Maria Curie-Sklodowska University \protect \\
20-031 Lublin, pl.~Marii Curie-Sklodowskiej 1, Poland}

%%%%%%%%%%%%%%%%%%%%%%%%%%%%%%%%%%%%%%%%%%%%%%%%%%%%%%%%%%%%%%%%%%%%
\date{\today}

%%%%%%%%%%%%%%%%%%%%%%%%%%%%%%%%%%%%%%%%%%%%%%%%%%%%%%%%%%%%%%%%%%%%%%%%%%%%%%%%%%%%%%%%%%%%%%%%%%%
\begin{abstract}
Based on the analytical technique the effect of the static magnetic field on the s-wave holographic 
superconductor with dark matter sector of  $U(1)$-gauge field type coupled to the Maxwell field 
has been examined. In the probe limit, we obtained the mean value of the condensation operator.
The nature of the condensate in an external magnetic field as well as the behavior of the 
critical field close to the transition temperature has been revealed. The obtained upturn of the
critical field curves as a function of temperature, both in four and five spacetime dimensions,  
is a fingerprint of the strong coupling approach.   
\end{abstract}
%%%%%%%%%%%%%%%%%%%%%%%%%%%%%%%%%%%%%%%%%%%%%%%%%%%%%%%%%%%%%%%%%%%%%%%%%%%%%%%%%%%%%%%%%%%%%%%%%%%%

\maketitle
%%%%%%%%%%%%%%%%%%%%%%%%%%%%%%%%%%%%%%%%%%%%%%%%%%%%%%%%%%%%%%%%%%%%%%%%%%%%%%%%%%%%%%%%%%%%%%%%%%%%%%%%
%%%%%%%%%%%%%%%%%%%%%%%%%%%%%%%%%%%%%%%%%%%%%%%%%%%%%%%%%%%%%%%%%%%%%%%%%%%%%%%%%%%%%%%%%%%%%%%%%%%%%

%%%%%%%%%%%%%%%%%%%%%%%%%%%%%%%%%%%%%%%%%%%%%%%%%%%%%%%%%%%%%%%%%%%%%%%%%%%%%%%%%%%%%%%%%%%%%%%%%%%%%%%
\section{Introduction}
The examination of holographic superconductors coupled to the electromagnetic and 
dark matter fields is of a great importance due to the fact that it
provides a strong coupling view of the condensation phenomenon,
as well as, it may reveal effects of dark matter on a condensing field which in turn could 
provide a clue to new methods of searching it. 
In this paper we shall use
the AdS/CFT correspondence to study the effect of a static magnetic field on the s-wave superconductor
when the matter fields will be described by the Abelian-Higgs sector coupled to Maxwell $U(1)$-gauge field.
The first studies of the backreacting s-wave holographic superconductor with dark matter sector, modelled by the
aforementioned fields were conducted by some of the present authors in Ref.\cite{nak14}.

The AdS/CFT correspondence \cite{mal} having its roots in string theory plays an essential role in examination 
of strongly correlated 
condensed matter systems. The correspondence in question binds string theory on asymptotically anti-de Sitter spacetime (AdS)
to a conformal field theory on the boundary. This fact is crucial for obtaining correlation functions 
in a strongly interacting field theory using a dual classical gravity attitude \cite{wit}. 

The successful implementations of this correspondence comprises the range from nuclear 
physics to condensed matter problems. 
The main advantage of using the string theory formalism for condensed matter problems  
stems from the fact that 
contrary to the bewildering complexity of the strong correlated systems, they can be treated 
as weakly coupled \cite{sachdev2012} being subject to perturbation
description. In this point of view, the temperature of the considered system is defined as the Hawking black hole 
temperature in the bulk.

Among many attempts to attack strongly coupled quantum problems by the gravity dual description  
the famous prediction of the universal 
ratio of the shear viscosity to the volume density of entropy has to be mentioned \cite{kovt05}. This finding 
explained some results of heavy ion collisions at the Relativistic  Heavy Ion Collider (RHIC) \cite{mat07}. The 
applications of the approach to condensed matter problems is overwhelming. 
In spite of the mentioned construction of the holographic s-wave superconductor \cite{har08}  
(for a recent review and updated references see \cite{reviews}), there exist 
models of d-wave \cite{d-wave}, p-wave \cite{p-wave}, two-band superconductors \cite{2bandsc} 
as well as examples of single 
band chiral superconducting states of spin singlet ($d+id^{'}$) \cite{chen2011} or triplet 
($p_x \pm ip_y$) \cite{chiral-p}.  
The Hall conductivity calculations for $d+id^{'}$ state, seem to  be  in the apparent 
conflict \cite{hongliu2013} with the weak coupling results \cite{rog-kiw2014}. 
%and of spin triplet $p_x\pm p_y$ character \cite{chiral-p}. 

Other interesting applications of the gauge-gravity duality to the condensed matter include the description 
of the strange metal \cite{sachdev2010},
the metal-insulator transition \cite{zaanen2014}, the metal-superconductor phase transition 
\cite{herzog2009,horowitz2011} or calculations of the
Hall and Nernst effects \cite{har07,har07a}. Some holographic results \cite{spintronics} may be of 
 importance for the emerging
branch of condensed matter physics called spintronics  \cite{dietl2014}, whose main aim is 
to replace existing electronic devices utilizing charge of electrons by novel system using spins of carriers 
to manipulate information.

Technically, the Abelian symmetry is broken outside  AdS black hole and results in the formation of the holographic 
superconductor as a charged scalar condensation in the dual CFT theory \cite{gub05}-\cite{gub08}.
The expectation value of charged operators undergoes the $U(1)$-symmetry breaking
and the second order phase transition comes into being \cite{har08,hor08,har08b} below characteristic temperature $T_c$. 
Moreover, the backreaction causes that even uncharged scalar fields may form a condensation in $(2+1)$-dimensions
\cite{har08b} in the apparent contradiction to the celebrated Mermin-Wagner theorem \cite{mermin1966}. 
It was also revealed that in the case of p-wave holographic models the phase transition 
leading to the formation of vector hair, changed from the second order to the first one.  This phenomenon depends 
on the strength of the gravitational coupling \cite{amm10}-\cite{cai11}.
%%%%%%%%%%%%%%%%%%%%%%%%%%%%%%%%%%%%%%%%%%%%%%%%%%%%%%%%%%%%%%%%%%%%%%%%%%%%%%%%

\par
The details of the approach and the obtained results depend on the gravity background, the kind of the 
electromagnetic theory and
the symmetry of the condensing field.  
%The other theories of gravity play also a significant role in holographic superconductivity studies.
In particular, it turned out that in Gauss-Bonnet (GB) gravity  the higher curvature corrections one considers 
the harder for 
the condensation to occur \cite{gre09}-\cite{kan11}. In the Einstein-Maxwell-dilaton gravity a generalized model built  
from the most extensive covariant gravity Lagrangian with at most two derivatives of fields \cite{apr10} was 
given \cite{liu10}. 
On the other hand, gravity theory including dilaton field was also investigated in the case of the potential 
models of superfluids and superconductors \cite{sal12}. Horava-Lifshitz gravity \cite{horava} 
and string/M theory \cite{mth} were also paid attention to. 

Different kinds of electro-magnetic theories were studied in the context of holographic superconductivity (see for 
example \cite{bi} and references therein).  For instance, the presence of the Born-Infeld scale parameter 
decreases the critical temperature and the ratio of the gap frequency in conductivity compared to the Maxwell 
electrodynamics. In fact, due to holographic superconductivity studies \cite{nonlin} 
the non-linear electrodynamics has acquired much attention recently. On the other hand, analytical 
methods of investigations of the holographic superconductivity with or without a backreaction were examined 
in Refs. \cite{anal}.
%%%%%%%%%%%%%%%%%%%%%%%%%%%%%%%%%%%%%%%%%%%%%%%%%%%%%%%%%%%%%%%%%%%%%%%%%%%%%%
\par
The other  subject especially interesting for us is the examination of magnetic field in the area of 
holographic superconductors.
In the paper \cite{alb08} it was shown that by adding magnetic charge to a black hole the holographic
superconductor could be immersed in the external magnetic field. On the other hand, when magnetic
fields increase the condensation shrinks in size in accordance with experiments. It was also observed that 
there exists a critical
value of magnetic field below which a charged condensate forms due to the second-order phase transition
\cite{nak08}. Studies in question also covered  the topic 
of critical magnetic fields and Abrikosov vortices \cite{prop}. Analytical studies of magnetic field influence
in the probe limit  with  the backreaction taken into account were conducted in GB gravity and in Born-Infeld, 
Weyl-corrected and non-linear electrodynamics \cite{ge10}-\cite{dey13}. It was revealed that the backreaction
caused the depression of the critical temperature value and it could enhance the upper critical magnetic field.
The non-linear parameter causes the decrease of the critical temperature and makes the condensation gap obtained 
within non-linear electrodynamics  greater than with the ordinary Maxwell one. It was also announced that 
GB coupling and Born Infeld parameter affect the critical magnetic field.
\par
The other tantalizing question which naturally arises is a request about possible matter configurations in 
AdS spacetime. It happens that strictly stationary Einstein-Maxwell spacetime with negative cosmological constant
does not allow for the existence of nontrivial configurations of complex scalar fields 
or form fields \cite{shi12}. The same situation takes place in the case of strictly stationary, simply connected 
Einstein-Maxwell-axion-dilaton spacetime with negative cosmological constant and arbitrary number of $U(1)$-gauge 
fields \cite{bak13}. 
%%%%%%%%%%%%%%%%%%%%%%%%%%%%%%%%%%%%%%%%%%%%%%%%
\par
The motivation for considering the dark matter sector comes from the astronomic observations and the quest for its
detection. In the late nineties the surveys for distant supernovae type Ia revealed that the high-redshifted
supernovae of this type appeared almost forty percent fainter (more distant) than expected \cite{accuniv}.
The other methods like analysis of cosmic microwave background \cite{ben03} or baryonic 
acoustic oscillations \cite{eis05} confirm that our Universe is filled with the negative pressure matter
triggering its acceleration. On the other hand, one has the strong evidence that almost 22 percent of the total energy
density of our Universe is in a form of a dark matter. We  only have vague ideas what it is made of.
The new model mimicking the dark matter sector was proposed (see, e.g., for the latest issues Refs.\cite{model}).
in which the standard model was coupled to the dark sector {\it via} an interactive term. Moreover, the aforementioned 
model was elaborated in the context of cosmic string interacting with dark string \cite{dark matter}.

The backreacting s-wave holographic superconductor with dark matter sector, where Maxwell field 
was coupled with Higgs field 
has been presented recently by two of the present authors \cite{nak14}. The main result of 
that paper was lowering of the critical 
temperature of the superconductor with backreaction parameter for the positive values 
of the effective coupling $\tilde{\alpha}$ to the dark 
matter ($\tilde{\alpha}> 0$) and the opposite behavior, called retrograde condensation, for the 
negative value of it. Present studies indicate that in order to have well defined 
 dimensionless expectation value of the condensation operator,
$\tilde{\alpha}$ should be limited to the positive values, which confines the value of the original coupling 
to be less than 2. It 
will be interesting to check if this behavior remains intact when the backreaction will be taken into account.

\par 
%%%%%%%%%%%%%%%%%%%%%%%%%%%%%%%%%%%%%%%%%%%%%%%%%%%%%%%%%
%On the other hand, the second ingredient of our research magnetic field s are indicated on scales ranging from stars, gala%ctic scales and far beyond.
The magnetic fields encountered in the Universe span scales ranging from stars to galaxies and far beyond. 
It plays a key role in astrophysical and cosmological phenomena (see, e.g., Ref.\cite{magnetic}).
It is hoped that the knowledge of the detailed behavior of the holographic superconductor in the background with dark matter
sector and in the presence of magnetic field might help not only to understand the behavior of strongly coupled superconductors
but also in elucidating the nature of the dark field itself and proposing new methods of its detection.

\par
The paper is organized as follows. In Sec. II we derive basic equations of the underlying system.
Sec. III will be devoted to studies of the dimensionless expectation value for the condensation operator.
In Sec. IV we consider the influence of magnetic field on the holographic superconductor with dark matter
sector, while Sec. V concludes our investigations.

%%%%%%%%%%%%%%%%%%%%%%%%%%%%%%%%%%%%%%%%%%%%%%%%%%%%%%%%%%%%%%%%%%%%%%%%%%%%%%%%%%%%%%%%%%%%%%%%%%%%%
\section{Action and equations of motion}
The motivation standing behind our investigations is to take into account
the model
in which {\it dark matter} sector is coupled to the Standard Model. It has its roots in astrophysical
observations of $511$ keV gamma rays detected by Integral/SPI \cite{integral} as well as
the experiments showing the electron positron excess in galaxy, revealed by ATIC/PAMELA \cite{atic, pamela}. Depending on
the conducted experiments their energies vary from a few GeV to a few TeV. These facts can be explained as annihilation
of dark matter into electrons. Namely, it may happen that below the GeV-energy scale the interaction term in the model
in question is in the form of a direct coupling between $U(1)$-gauge strength field representing
{\it dark matter} sector and Maxwell strength tensor.

On the other hand,  the other possible evidence revealing new physics
is the $3.6\sigma$ discrepancy between measured value of the muon anomalous magnetic moment 
and its prediction in the Standard Model \cite{muon}.
From the cosmological point of view, these kinds of dark matter models admitted cosmic string like solutions,
topological defects \cite{hartmann2014} which might arise in the early Universe \cite{dark matter}.
Having all these facts in mind, it will be not amiss to search for the imprints
of {\it dark matter} sector in holographic superconductor physics, which can be potentially observed  
in the future experiments.
\par
The gravitational part of the model action implies
\be
S_{g} = \int \sqrt{-g}~ d^n x~  \frac{1}{2 \kappa^2}\bigg( R - 2\Lambda\bigg), 
\ee
where $\kappa^2 = 8 \pi G_{n}$ is an n-dimensional gravitational constant.
The cosmological constant will be given by
$\Lambda = - \frac{(n-1)(n-2)}{2L^2}$, 
where $L$ is the radius of the AdS spacetime.
%%%%%%%%%%%%%%%%%%%%%%%%%%%%%%%%%%%%%%%%%%%%%%%%%%%%%%%%%%%%%%%%%%%%%%
Our aim is to study the s-wave superconductivity in the presence of magnetic field. To describe it a 
complex charged scalar field $\psi$
will be introduced into the matter action.
%%%%%%%%%%%%%%%%%%%%%%%%%%%%%%%%%%%%%%%%%%%%%%%%%%%%%%%%
Moreover, the Abelian-Higgs sector is coupled to the second $U(1)$-gauge field.
It is provided by the following action:
\be
\label{s_matter}
S_{m} = \int \sqrt{-g}~ d^nx  \bigg( 
- \frac{1}{4}F_{\mu \nu} F^{\mu \nu} - \left [ \nabla_{\mu} \psi - 
i q A_{\mu} \psi \right ]^{\dagger} \left [ \nabla^{\mu} \psi - i q A^{\mu} \psi  \right ]
- V(\psi) - \frac{1}{4} B_{\mu \nu} B^{\mu \nu} - \frac{\alpha}{4} F_{\mu \nu} B^{\mu \nu}
\bigg), 
\ee  
where the scalar field potential satisfies
$V(\psi) = m^2 |\psi|^2 + \frac{\lambda_{\psi}}{4} |\psi|^4$.
$F_{\mu \nu} = 2 \nabla_{[ \mu} A_{\nu ]}$
stands for the ordinary Maxwell field strength tensor, while
the second $U(1)$-gauge field $B_{\mu \nu}$ is given by  
$B_{\mu \nu} = 2 \nabla_{[ \mu} B_{\nu ]}$. Moreover, $m,~ \lambda_{\psi},~ q$ represent
mass, a coupling constant and charge related to the scalar field $\psi$, respectively.
On the other hand, $\alpha$ is a coupling constant between $U(1)$ fields.
In order to be compatible with the current observations it should be on the order of $10^{-3}$.

%%%%%%%%%%%%%%%%%%%%%%%%%%%%%%%%%%%%%%%%%%%%%%%%%%%%%%%%%%%%%%%%%%%%%%%%%%%%%%%%%%%%%%%%%%%%%%%%%%%%%

In order to proceed further one introduces a general line element provided by 
\be
\label{metric}
ds^2 = - f(r)~dt^2 + \frac{dr^2}{f(r)} + r^2 h_{i j} dx^i dx^j,
\ee 
where $f$ is a function of  $r$-coordinate, $h_{i j}$ is the metric tensor on the 
$(n-2)$-dimensional submanifold. In the case under consideration  
we take into account a planar Schwartzchild black hole, for which metric function $f(r)$ yields
\be
f(r) = \frac{r^2}{L^2} - \frac{r_{+}^{n-1}}{L^2 r^{n-3}},
\ee
Without loss of generality we put $L = 1$. 
We assume also that the considered gauge fields posses only the temporal components which also 
depend only on the radial coordinate $r$, i.e., $A_{t} = \phi(r),~B_{t} = \eta(r)$.\\
Equations of motion for the underlying system (see Ref.\cite{nak14}) for the explicit forms and derivations)
can be rewritten in a more convenient variable 
$z = \frac{r_{+}}{r}$. They are given by 
the following set of differential equations:
\ben
%%%%%%%%%%%%%%%%%%%%%%%%%%%%%%%
\label{eq-phi}
\partial^{2}_{z} \phi &-& \bigg(  \frac{n-4}{z} \bigg)~\partial_{z} \phi 
- \frac{2 r_{+}^2}{ \tilde{\alpha} z^4} \frac{q^2 \psi^2}{f}\phi = 0, \\
%%%%%%%%%%%%%%%%%%%%%%%%%%%%%%%%%%%%%%
\label{eq-psi}
\partial^{2}_{z} \psi &-& \bigg(  \frac{n-4}{z} - \frac{1}{f} \partial_{z}f  \bigg)~\partial_{z} \psi 
+ \frac{r_{+}^2 ~q^2 ~\phi^2}{f^2 z^4} \psi
- \frac{r_{+}^2}{z^4} \frac{1}{2f} \frac{ \partial V(\psi)}{\partial \psi}  =0,\\
%%%%%%%%%%%%%%%%%%%%%%%%%
\label{eq-eta}
\partial_{z} \eta &=& { c_{1} \over r_{+}^{n-3}} z^{n-4} - \frac{ \alpha}{2} \partial_{z} \phi,
\een
where we introduce $\tilde{\alpha} = 1 - \frac{\alpha^2}{4}$ and the integration constant $c_{1}$.

%%%%%%%%%%%%%%%%%%%%%%%%%%%%%%%%%%%%%%%%%%%%%%%%%%%%%%%%%%%%%%%%%%%%%%%%%%%%%%%%%%%%%%%%%%%%%%%
\section{Holographic superconductor with dark matter sector}
In the holographic approach to superconductivity one is interested in the temperature below which 
the scalar charged field $\psi$
condensates. Above the critical temperature the normal metal solution $\psi(z) = 0$ is realized. The spontaneous 
symmetry breaking - hallmark of the superconductivity, occurs 
if a charged operator acquires non-zero vacuum expectation value. Charge operator is dual to the field $\psi$ and 
its expectation value is deduced \cite{reviews}
from the behaviour of the scalar field near the boundary $z \rightarrow 0$. This introductory section shows how the 
coupling constant between matter and dark matter 
sectors affects the superconducting temperature. In the next section which constitutes the main part of the paper, we study the 
influence of $\alpha$ on the critical magnetic field.
\par
Here, we shall apply matching method \cite{gre09} in order to obtain the quantities
we are in search of. To commence with, one finds
the leading order solutions near the black hole event horizon and in the asymptotically
AdS region. Then one matches them smoothly at the intermediate point $z_m$. We consider
the range of $z$-coordinate from $1 \ge z \ge 0$ and $1 > z_m > 0$.
\par
To begin with one specifies the boundary conditions, i.e., in the asymptotically AdS region where
$z \rightarrow 0$ the solutions are provided 
\ben
\label{eq-phi-z0}
\phi &=& \mu - \frac{\rho}{r_{+}^{n-3}}z^{n-3}, \\
\label{eq-psi-z0}
\psi &=& C_{-}z^{\Delta_{-}} + C_{+}z^{\Delta_{+}},
\een
and the regularity conditions at the black hole event horizon, $z =1$, are given by 
\ben
\phi(1) &=& 0, \\
\p_{z}f~ \p_{z} \psi &=& \frac{1}{2} \p_{\psi} V~r_{+}^2.
\een
In the above relations $C_{\pm}$ are expectation values of the operators we are looking for and 
$\Delta_{\pm} = \frac{1}{2} [  n-1 \pm \sqrt{(n-1)^2 + 4m^2} ]$.
\par
Next, expanding $\phi$ and $\psi$ functions in the Taylor series near $z=1$ and using
equations of motion near the aforementioned point, one can readily find that
the relations describing $\phi$ and $\psi$ yield 
\ben
\label{eq-phi-z1}
\phi(z) &=& - \partial_{z} \phi_{|z=1} (1 - z) + \frac{1}{2}
\left \{  (n-4) + \frac{2 r_{+}^2 q^2}{ \partial_{z} f_{|z=1} \tilde{\alpha}} \psi(1)^2   \right \} 
\partial_{z} \phi_{|z=1} (1 - z)^2  + \dots, \\
%%%%%%%%%%%%%%%%%%%%%%%%%%%%%%%%%%%%%%%%%%%%%%%
\label{eq-psi-z1}
\psi(z) &=& \psi(1) -
r_{+}^2 \frac{m^2}{\partial_{z} f_{| z = 1}} \psi(1) (1 - z) + \nonumber \\ 
&+& \frac{1}{4}
\left \{ -r_{+}^2 q^2 \frac{(\partial_{z} \phi_{|z=1})^2}{ (\partial_{z} f_{|z=1})^2}
+ \frac{ r_{+}^2}{ \partial_{z} f_{|z=1}} m^2 
[ n-8- \frac{1}{ \partial_{z} f_{|z=1} } \left ( \partial_{z}^2 f_{|z=1} - r_{+}^2 m^2 \right )  ]
\right \} \psi(1) (1-z)^2 + \dots
\een
In order to match these sets of asymptotic solutions at the intermediate point $z_m$, one should
satisfy the following conditions for functions and their derivatives. The system of the
equations in question imply
\ben
\label{eqf-1}
C_{+} z_{m}^{\Delta_{+}} &=& 
\psi(1) -
r_{+}^2 \frac{m^2}{f^{'}(1)} \psi(1) (1 - z_{m}) + \nonumber \\ 
&+& \frac{1}{4}
\left \{ - r_{+}^2 q^2 \frac{\phi^{'}(1)^2}{ f^{'}(1)^2}
+ \frac{ r_{+}^2}{ f^{'}(1)} m^2 
[ n-8 - \frac{1}{ f^{'}(1) } \left ( f^{''}(1) - r_{+}^2 m^2 \right )  ]
\right \} \psi(1) (1-z_{m})^2 ,  \\
%%%%%%%%%%%%%%%%%%%%%%%%%%%%%%%%%%%%%
\label{eqf-2}
\Delta_{+} C_{+} z_{m}^{\Delta_{+} -1} &=&
r_{+}^2 \frac{m^2}{f^{'}(1)} \psi(1) + \nonumber \\ 
&-& \frac{1}{2}
\left \{ - r_{+}^2 q^2 \frac{\phi^{'}(1)^2}{ f^{'}(1)^2}
+ \frac{ r_{+}^2}{ f^{'}(1)} m^2 
[ n-8 - \frac{1}{ f^{'}(1) } \left ( f^{''}(1) - r_{+}^2 m^2 \right ) ]
\right \} \psi(1) (1-z_{m}) , \\
%%%%%%%%%%%%%%%%%%%%%%%%%%%%%%%%%%%%%
\label{eqf-3}
\mu - \frac{\rho}{r_{+}^{n-3}}z_{m}^{n-3} &=&
- \phi^{'}(1) (1 - z_{m}) + \frac{1}{2}
\left \{  (n-4) + \frac{2 r_{+}^2 q^2}{ f^{'}(1) \tilde{\alpha}} \psi(1)^2   \right \} \phi^{'}(1) (1 - z_{m})^2 , \\
%%%%%%%%%%%%%%%%%%%%%%%%%%%%%%%%%%%%%
\label{eqf-4}
-(n-3)\frac{\rho}{r_{+}^{n-3}}z_{m}^{n-4} &=&
\phi^{'}(1)  -
\left \{  (n-4) + \frac{2 r_{+}^2 q^2}{ f^{'}(1) \tilde{\alpha}} \psi(1)^2   \right \} \phi^{'}(1) (1 - z_{m}),
\een 
where for brevity we used a prime to denote differentiation with respect to $z$-coordinate.
%%%%%%%%%%%%%%%%%%%
By virtue of the above system of differential equations we are looking for the quantities 
of our interest, i.e., $\psi(1),~ C_{+}$ and $\phi^{'}(1)$.
As far as the other parameters appearing in the above relations are concerned, one 
exchanges $r_{+}$ and $\rho$ for the black hole temperature and the critical temperature, respectively
and assumes that the rest of them are specified by the theory at hand.
%%%%%%%%%%%%%%%%%%%%%%%
First, from relation (\ref{eqf-4}) we find that $\psi(1)$ is given by the following equation 
\be
\psi(1)^2 = - \frac{1 - (1-z_{m})(n-4)}{2q^2 (1-z_{m})} \tilde{\alpha}(n-1) + 
\frac{1}{r_{+}^{n-2}} (n-3) \rho z_{m}^{n-4} \frac{\tilde{\alpha} (n-1)}{2 q^2 (1 - z_{m}) \beta},
\label{bbb}
\ee
where we set $\beta = - \frac{ \phi^{'}(1)}{r_{+}}$. Consequently, having in mind that $f^{'}(1) = - r_{+}^2 (n-1)$
as well as the definition of Hawking temperature of 
the black hole $T_{BH} = \frac{r_{+}}{4 \pi} (n-1)$, the above equation yields 
\be
\psi(1)^2 = \frac{ \tilde{\alpha} (n-1) [ 1 - (1-z_{m})(n-4)]}{2 q^2 (1-z_m)} 
\left ( \frac{T_{c}}{T_{BH}} \right )^{n-2} 
\left [ 1 - \left (  \frac{T_{BH}}{T_{c}} \right )^{n-2} \right ],
\ee
where the critical temperature implies the relation
\be
\label{Tc}
T_{c} = \frac{n-1}{4 \pi} 
\left [ \frac{(n-3) z_{m}^{n-4}}{ [1 - (1-z_{m})(n-4)]\beta} \right ]^{\frac{1}{n-2}} \rho^{\frac{1}{n-2}}.
\ee
Returning to the equations (\ref{eqf-1}) and (\ref{eqf-2}), we find that $C_{+}$ and $\beta$ are provided by
%%%%%%%%%%%%%%%%%%%%%%%%%%%%%%%%%%%%%%%%%%%%
\ben
\label{beta2}
\beta^2 &=& 
\frac{2\Delta_{+}(n-1)[2(n-1) + m^2(1-z_{m})]}{q^2 (1-z_{m})(\Delta_{+}(1-z_{m}) + 2z_{m})} +
m^2 \frac{2(n-1)(2-z_m) + m^2(1-z_{m})}{q^2 (1 - z_{m})}
, \\
%%%%%%%%%%%%%%%%%%%%%%%%%%%%%%%%%%%%%%%%%%%%%%%%%
C_{+} &=& 
\frac{  z_{m}^{1-\Delta_{+}} [ m^2 ( 1 -z_{m}) + 2 (n-1) ] }{ \Delta_{+} (n-1) ( 1 - z_{m} + \frac{2 z_{m}}{\Delta_{+}} ) } \psi(1).
\een

%%%%%%%%%%%%%%%%%%%%%%%%%%%%%%%%%%%%%%
%%%%%%%%%%%%%%%%%%%%%%%%%%%%%%%%%%%%%%%%%%%%%%%%%%%%%%%%%%%%%%%%%%%%%%%%%%%%%%%%%%%%%%%%%%%%%%%%%%%%%%5
To proceed further, we recall that an asymptotic behavior of the scalar field may be cast in the form as
\be
\psi \sim \frac{<O_{+}>}{r^{\Delta_{+}}} = \frac{<O_{+}>}{r_{+}^{\Delta_{+}}} z^{\Delta_{+}} = C_{+} z^{\Delta_{+}}. 
\ee
Having in mind the above relation and using 
the definition of the black hole temperature, it can be found that the expectation value of the condensation
operator is provided by 
\be
<O_{+}> =\bigg( \frac{4\pi}{n-1} \bigg)^{\Delta_{+}}~ T_{BH}^{\Delta_{+}}~ C_{+},
\ee
while the dimensionless expectation value yields
\ben
\label{expectation_value}
\frac{<O_{+}>}{T_{c}^{\Delta_{+}}} &=&
\bigg( \frac{4\pi}{n-1} \bigg)^{\Delta_{+}} 
\frac{  z_{m}^{1-\Delta_{+}} [ m^2 ( 1 -z_{m}) + 2 (n-1) ] }
{ \Delta_{+} (n-1) ( 1 - z_{m} + \frac{2 z_{m}}{\Delta_{+}} ) } 
\sqrt{\frac{ \tilde{\alpha} (n-1) [ 1 - (1-z_{m})(n-4)]}{2 q^2 (1-z_m)} } \nonumber \\
& \times &
\left ( \frac{T_{BH}}{T_{c}} \right )^{\Delta_{+} - \frac{n-2}{2}} 
\sqrt{ 1 - \left (  \frac{T_{BH}}{T_{c}} \right )^{n-2} }.
\een
It can be readily found, that near the critical temperature, $T_{BH}/T_c \rightarrow 1$, the 
expectation value of the operator in question is of the 
standard form $<O_{+}> \sim \sqrt{1 - \frac{T_{BH}}{T_{c}} }$. This behaviour is a typical mean field
theory result describing a second order phase transition. The prefactor, however, depends on the coupling
constant $\tilde{\alpha}$ in a non-analytic way. This fact limits the bare coupling values to $0<\alpha<2$.

%%%%%%%%%%%%%%%%%%%%%%%%%%%%%%%%%%%%%%%%%%%%%%%%%%%%%%%%%%%%%%%%%%%%%%%%%%%%%%%%%%%%%%%%%%%%%%%%%%%%%%%%%%%%%%%%%%%
%%%%%%%%%%%%%%%%%%%%%%%%%%%%%%%%%%%%%%%%%%%%%%%%%%%%%%%%%%%%%%%%%%%%%%%%%%%%%%%%%%%%%%%%%%%%%%%%%%%%%%%%%%%%%%%%%%%
\section{Magnetic field effect on dark matter sector holographic superconductor}
Because of the fact that in the dark matter model one has to do with 
two $U(1)$-gauge fields, we may introduce the magnetic field in twofold way.
Namely it can be supposed that $A = \phi~ dt + \tilde{B}~x~dy$ or $B = \eta~ dt + \tilde{B}~x~dy$.
Since the scalar field is uncharged with respect to the $B_{\mu}$ field, the first method 
of introducing magnetic field reduces our investigations to only one gauge field. Far more interesting 
situation one gets when we allow magnetic part of the $A_{\mu}$ field
to be induced by the kinetic mixing with the $B_{\mu}$. This correspond to the second of the aforementioned methods.
\par 
The effect of the magnetic field will be examined by applying the same procedure as in the preceding section. Namely, 
we shall find the asymptotic value of the scalar field close to the boundary, but now in the presence of the 
magnetic field.
%%%%%%%%%%%%%%%%%%%
To begin with we solve the equation for the $A_{y}$ component. Further,
assuming that this component depends only on $x$ coordinate we obtain
\be
r^{-4} \partial_{x} \bigg(
\partial_{x} B_{y} + \frac{\alpha}{2} \partial_{x} A_{y}
\bigg) = 0.
\ee
Consequently, for $B_{y} = \tilde{B}~ x$, the solution implies the following behavior of $A_y$ 
\be
A_{y} = \frac{2}{\alpha}~( c_{0} - \tilde{B})~x + c_{1},
\ee
where $c_{0}$ and $c_{1}$ are integration constants.
Consider now the relation describing the scalar field $\psi$
\be \label{pp}
\partial_{r}^2 \psi + 
\bigg( \frac{n-2}{r}  + \frac{\partial_{r} f}{f}  \bigg) \partial_{r} \psi
- \frac{m^2}{f} \psi + \frac{q^2 \phi^2}{f^2} \psi +
\frac{1}{f r^2} \bigg(
\partial_{x}^2 \psi -q^2 ~( \frac{2}{\alpha}(c_{0} - \tilde{B})x +c_{1} )^2 \psi
\bigg) = 0.
\ee
In order to solve the above equation (\ref{pp}), we set $\psi$ in the separable form
as $F(r) X(x)$. It enables us to find that $X$ is subject to the relation
\be
\partial_{x}^2 X - q^2~ ( \frac{2}{\alpha}(c_{0} - \tilde{B})x +c_{1} )^2 X = -l^2 X,
\ee 
where $l^2$ is a separation constant. 
%%%%%%%%%%%%%%%%%%%%%%%%%%%%%%%%%%%%%%%%%%%
We will be interested in the solution that is regular everywhere and decays as the $x$ tends to $\pm \infty$.
The above equation  can be cast in the form 
\be
\label{parab_cylind_def}
Y^{''} + \left ( \nu + \frac{1}{2} - \frac{1}{4}y^2 \right )Y = 0,
\ee
where $'$ means the derivative with respect to $y$-coordinate. The solutions of this equation 
with the aforementioned boundary condition are parabolic cylinder functions $D(\nu;y)$.
By virtue of the substitution $y = 2 \frac{ \sqrt{ q ( \tilde{B} - c_0 ) } }
{ \sqrt{ \alpha} } x - \frac{ c_1 \sqrt{ q \alpha} }{ \sqrt{ \tilde{B} - c_0 } }$ we may rewrite our equation for $X$
as 
\be
\frac{ d^2}{ d y^2} X + \left ( \frac{1}{4} \frac{ \alpha l^2}{ q (\tilde{B} - c_0)} - \frac{1}{4} y^2 \right ) X = 0.
\ee
%%%%%%%%%%%%%%%%%%%%%%%%%%%%%%%%%%%%%%%%%%%%%%%%%%%%%%%%%%%%%%%%%%%%%%%%%%%%%%%%%%

Comparison of the above relation with (\ref{parab_cylind_def}) allows us to identify $\nu$ as 
$\nu = \frac{ \alpha l^2 - 2q ( \tilde{B} - c_0 ) }{ 4q ( \tilde{B} - c_0) }$.
In the case under consideration the explicit solution of $X$ is given by the $D(\nu;x)$ function as 
\be
X = \tilde{c}~ D\bigg( \frac{ - 2\tilde{B}q + 2 c_{0}q + l^2 \alpha}{4q(\tilde{B} - c_{0})};~ 
\frac{2 \sqrt{q}~x \sqrt{\tilde{B} - c_{0}}}{\sqrt{\alpha}} - \frac{c_{1}\sqrt{\alpha q}}{\sqrt{\tilde{B} 
- c_{0}}} \bigg),
\ee
where $\tilde{c}$ is an integration constant.
Moreover, demanding that $\nu = 2k$, where $k= 0,1,2,\dots$, we find
\be
l^2 = (4k + 1) \frac{2 \bar{B} q}{\alpha},
\ee 
where we have defined $\bar{B} = \tilde{B} - c_{0}$.
%%%%%%%%%%%%%%%%%%%%%%%%%%%%%%%%%%%%%%%%%%%%%%%%
For this particular form of the separation constant, our solution reduces to the product 
of the Gaussian function and the Hermite polynomial (the so-called modified Hermite polynomial). 
It is everywhere regular, possesses $k$ nodes and decays as the argument goes to $\pm \infty$.

%%%%%%%%%%%%%%%%%%%%%%%%%%%%%%%%%%%%%%%%%%%%%%%
Consequently using the above relation we arrive at the  expression
\ben
\partial_{r}^2 F + \frac{n-2}{r} \partial_{r}F + \frac{q^2 \phi^2}{f^2} F +
\frac{1}{f} \bigg( 
\partial_{r} f \partial_{r} F - m^2 F - \frac{1}{r^2} (4k+1) \frac{2 \bar{B} q}{\alpha} F
\bigg) = 0.
\een
In the next step, we change variable for $z$-coordinates. It yields
\ben
\partial_{z}^2 F - \frac{n-4}{z} \partial_{z}F + \frac{r_{+}^2 q^2 \phi^2}{ f^2 z^4} F +
\frac{1}{z^4 f} \bigg(
z^4 \partial_{z}f \partial_{z}F -  r_{+}^2 m^2 F - z^2 (4k+1)\frac{2 \bar{B} q}{\alpha} F 
\bigg) = 0
\een
The approximate solution of the above equation
can be achieved by the matching method proposed in Ref.\cite{gre09}, using approximations of the 
solutions by truncated series. Namely,
having in mind the regularity condition in $z = 1$, we obtain the following:
\be
\left [ 
z^4 \partial_{z} f \partial_{z}F - m^2 r_{+}^2 F - z^2 (4k+1)\frac{2 \bar{B}q}{\alpha} F 
\right ]_{| z = 1} = 0.
\ee
Then, it can be readily shown that 
\be
\label{reg-mag}
(\partial_{z}F)_{z = 1} = \frac{1}{(\partial_{z} f)_{|z=1}}
\left [ 
m^2 r_{+}^2 F(1) + (4k+1)\frac{2 \bar{B}q}{\alpha} F(1)
\right ].
\ee
Evaluating the equation of motion for scalar field at $z=1$, one gets the relation
%%%%%%%%%%%%%%%%%%%%%%%5555
%%%%%%%%%%%%%%%%%%%%%%%%%%%%%%%%%%%%%%%%%%%5
\ben
2 (\partial_{z}^2 F)_{z=1} &=& F(1) \bigg \{ - \frac{n-4}{n-1} \left [
m^2 + (4k+1) \frac{2 \bar{B} q}{r_{+}^2 \alpha}
\right ]  + \nonumber \\
&-& \left [
(n-2) + \frac{1}{n-1} \left (m^2 + (4k+1) \frac{2 \bar{B} q}{r_{+}^2 \alpha} \right )
\right ] \left ( \frac{-1}{n-1} \right ) \left ( m^2 + (4k+1) \frac{2 \bar{B} q}{r_{+}^2 \alpha} \right )
+ \nonumber \\
&-& \left ( \frac{2}{n-1} \right ) \left ( m^2 + (4k+1) \frac{2 \bar{B} q}{r_{+}^2 \alpha} \right )
+ \nonumber \\
&-&  \frac{q^2}{(n-1)^2} 
\left ( - \frac{ (\partial_{z} \phi)_{|z=1}}{ r_{+}} \right )^2  
+ \frac{2m^2}{n-1}  \bigg \}, 
\een
%%%%%%%%%%%%%%%%%%%%%%%%%%%%%%%%%%%%%%%%%%5  
where we have used Eq.(\ref{reg-mag}) to eliminate $(\partial_{z}F)_{|z=1}$ 
and set 
$$(\partial_{z} f)_{|z=1} = - r_{+}^2 (n-1), \qquad 
(\partial_{z}^2 f)_{|z=1} = r_{+}^2~(n-1)(6-n).$$
Near $z=1$, the Taylor expansion of $F(z)$ provides us the relation
\be
F(z) = F(1) - (1-z) (\partial_{z}F)_{|z=1} + \frac{1}{2}(1 - z)^2 (\partial_{z}^2 F)_{|z=1} + \dots.
\ee
Next, matching this series solution to the asymptotic solution at some intermediate point, one arrives at
%%%%%%%%%%%%%%%%%%%%%%%%%%
\ben
\label{B1}
D_{+} z_{m}^{\Delta_{+}} &=& F(1) \bigg \{
1 + (1 - z_{m})A + \frac{1}{4} (1 - z_{m})^2 \bigg [  A^2  - c \bigg]
\bigg \} \\
%%%%%%%%%%%%%%%%%%%%%%%%%%%%%%%%5
\label{B2}
\Delta_{+} D_{+} z_{m}^{\Delta_{+} - 1} &=&
F(1) \bigg \{
- A - \frac{1}{2} (1 - z_{m})\bigg [  A^2  - c \bigg]
\bigg \},
\een
where for  convenience we introduce the quantities
\ben
A = \frac{1}{n-1} \bigg [ m^2 + (4k+1) \frac{2q \bar{B}}{\alpha r_{+}^2} \bigg ], \\
%b = 10 - n - \frac{6 - (n-3)(n-4)}{n-1}, \\
c =  - \frac{2 m^2}{n-1} + \frac{q^2}{n-1} \left ( - \frac{ (\partial_{z} \phi)_{|z=1}}{r_{+}} \right )^2_{| \bar{B} \neq 0}.
\een

%%%%%%%%%%%%%%%%%%%%%%%%%%%%%%%%%%%%%%%%%%%%%%%%%%%%%%%%%%%
We remark that $D_{+}$ plays the analogous role as $C_{+}$ in the zero magnetic field case. However, now we are 
interested in finding the critical magnetic field, which means that the behaviour of the condensate is fixed (the condensate is almost vanishing, $\psi^2 \sim 0$) and
the only relevant quantity that one wants to find is $\bar{B}$.  
%%%%%%%%%%%%%%%%%%%%%%%%%%%%%%%%%%%%%%%  
As far as the equation (\ref{eq-phi}) for the electric component of gauge field is concerned,
we may rewrite it (after discarding $\psi^2$ term) in the form 
\be
\partial_{z}^2 \phi - {(n-4)\over z} \partial_{z} \phi =0.
\ee
On the other hand, the Taylor series solution enables us to find that
\be
\phi(z) = - (\partial_{z} \phi)_{|z=1} (1 - z) + \frac{1}{2}(1-z)^2 (\partial^2_{z} \phi)_{|z=1},
\ee 
where we have used fact that at the black hole event horizon one has that $\phi(1) = 0$.
%%%%%%%%%%%%%%%%%%%%%%%%%%%%%%%%%%%%%%%%%%%%%%%%%%%%%

Matching this with the asymptotic expression enables us to write
\be
\bigg(- \frac{ (\partial_{z} \phi )_{|z=1}}{r_{+}} \bigg)_{| \bar{B} \neq 0} = 
\frac{(n-3) z_{m}^{n-4}}{1 - (n-4)(1-z_{m})} \frac{\rho}{r_{+}^{n-2}}. 
\ee
Consequently, having in mind that the charge density $\rho$ 
can be expressed by the critical temperature at zero magnetic field Eq.(\ref{Tc})
and using the definition of the black hole temperature, we arrive at
\be
\bigg(- \frac{ (\partial_{z} \phi )_{|z=1}}{r_{+}} \bigg)_{| \bar{B} \neq 0}^2 = 
\beta^2 \bigg ( \frac{T_{c}}{T_{BH}}\bigg)^{2(n-2)},
\ee
where $\beta$ is given by the equation (\ref{beta2}).
%%%%%%%%%%%%%%%%%%%%%%%%%%%%%%%%%%%%%%%%%%%%%%%%%%%%%%%%%%%%%%%%%%%%%%%%%%%%%%%%%%%%%
Let us remark, that the magnetic field introduces the additional dependence of 
$\beta = - \frac{ \phi^{'}(1)}{r_{+}}$ on the temperature. Namely, for $T_{BH} = T_c$ the quantity
in question reduces to $\beta$ defined in the equation (\ref{bbb}). This agrees with the fact that at
$T = T_c$, the critical magnetic field $\bar{B} = 0$, which is in accordance with experiments.
%%%%%%%%%%%%%%%%%%%%%%%%%%%%%%%%%%%%%%%%%%%%%%%%%%%%%%%%%5
Inserting equation (\ref{B1}) into (\ref{B2}) and solving for $A$, one has that
%%%%%%%%%%%%%%%%%%%%%%%%%
\ben
A &=& - \frac{  1 + \frac{\Delta_{+}}{z_m}(1-z_m) }{ (1-z_m)(1 + \frac{\Delta_{+}}{2z_m}(1-z_m) )  }
\bigg \{
1 + \nonumber \\
&\mp& \sqrt{1 + 
2 \frac{
 (1 - z_m)(1 + \frac{\Delta_{+}}{2 z_m}(1 - z_m) ) 
\left [ \frac{1}{2} c (1 - z_m) ( 1 + \frac{\Delta}{2z_m}(1 - z_m) - \frac{\Delta_{+}}{z_m} )  
\right ]
}
{ \left ( 1 + \frac{\Delta_+}{z_m}(1 - z_m)  \right )^2}
}
\bigg \}
\een
From the above relations we finally obtain the critical magnetic field strength we are looking for
\ben
\label{magnetic_field}
\bar{B} &=& \frac{\alpha (4 \pi)^2}{2 q (n-1)(4k+1)} T_{BH}^2 
\bigg \{
- \frac{m^2}{n-1} + \nonumber \\
&-& \tilde{b} \left [ 1 \mp 
\sqrt{1 + 
\frac{ \bar{b}^2
\left [
- \frac{2m^2}{n-1} + \frac{q^2}{(n-1)^2} \beta^2 \left ( \frac{T_{c}}{T_{BH}} \right )^{2(n-2)}
\right ] - \frac{2 \Delta_+}{z_m} \bar{b}  }
{ \left ( 1 + \frac{\Delta_+}{z_m}(1 - z_m)  \right )^2}
}
\right ]
\een 
where we set $\tilde{b}$ and $\bar{b}$ in the form as
\be
\tilde{b} = \frac{  1 + \frac{\Delta_{+}}{z_m}(1-z_m) }{ (1-z_m)(1 + \frac{\Delta_{+}}{2z_m}(1-z_m) )}, \qquad
\bar{b} = (1 - z_m) \left [ 1 + \frac{\Delta_{+}}{2 z_m}(1 - z_m) \right ].
\ee
%%%%%%%%%%%%%%%%%%%%%%%
The expression for $\bar{B}$ contains an arbitrary sign in front of the square root. We choose it in such a way that $\bar{B}$ is always positive. This amounts to choosing '$-$' for 
all values of other parameters that we considered.

%%%%%%%%%%%%%%%%%%%%%%%%%%%%%%%%%%%%%%%%%%%%%%%%%%%%%%%%%%%%%%%%%%%%%%%%
%%%%%%%%%%%%%%%%%%%%%%%%%%%%%%%%%%%%%%%%%%%%%%%%%%%%%%%%%%%%%%%%%%%%%%%
\section{Results and conclusions}
In our considerations we used the midpoint method to obtain the expectation value
of scalar operator which in turn represents the order parameter in s-wave holographic
superconductor. In order to fix an arbitrary parameter $z_m$, one takes into account the critical
temperature $T_{c}$ found in Ref.(\cite{nak14}) by means of the Sturm-Liouville method and 
compares it with the formula (\ref{Tc}). Of course, one ought to remember that the method in question
is valid in the vicinity of the critical temperature.

\begin{figure}[tbh]
\includegraphics[scale=1]{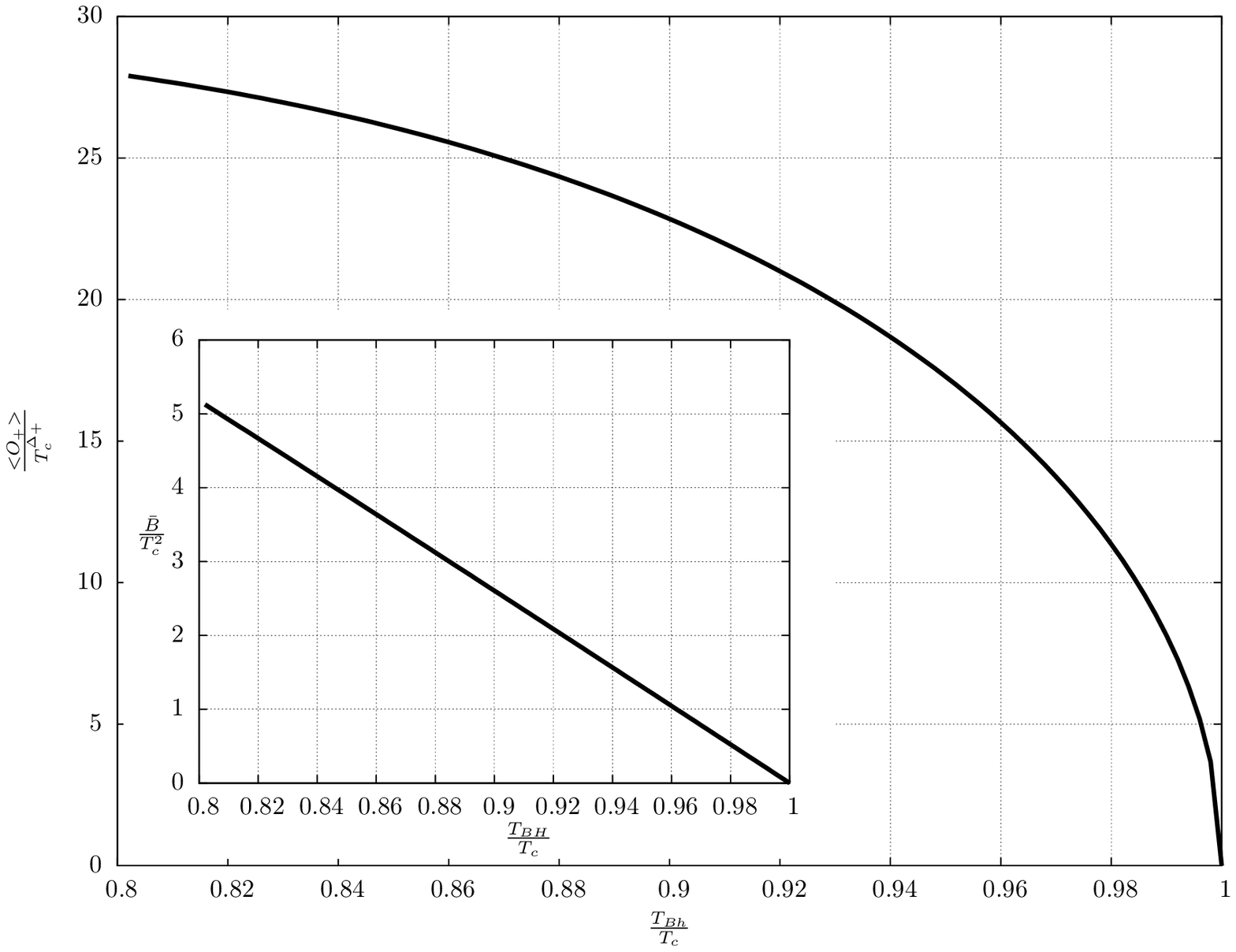}             
\caption{Normalized expectation value of operator $<O_{+}>$ and critical magnetic field as the functions of the 
dimensionless temperature. Spacetime dimension is $n=4$ and matching point $z_{m} = 0.344993$. 
Remaining parameters of the scalar field are $q = 1, ~~ m^2 = -2, ~~ \Delta_{+} = 2$. Gauge field
parameters are $k = 0$ and $\alpha = 0.5$. The insert shows the behavior of the critical magnetic field
close to $T_c$.}
\label{fig1}
\end{figure}

\begin{figure}[tbh]
\includegraphics[scale=1]{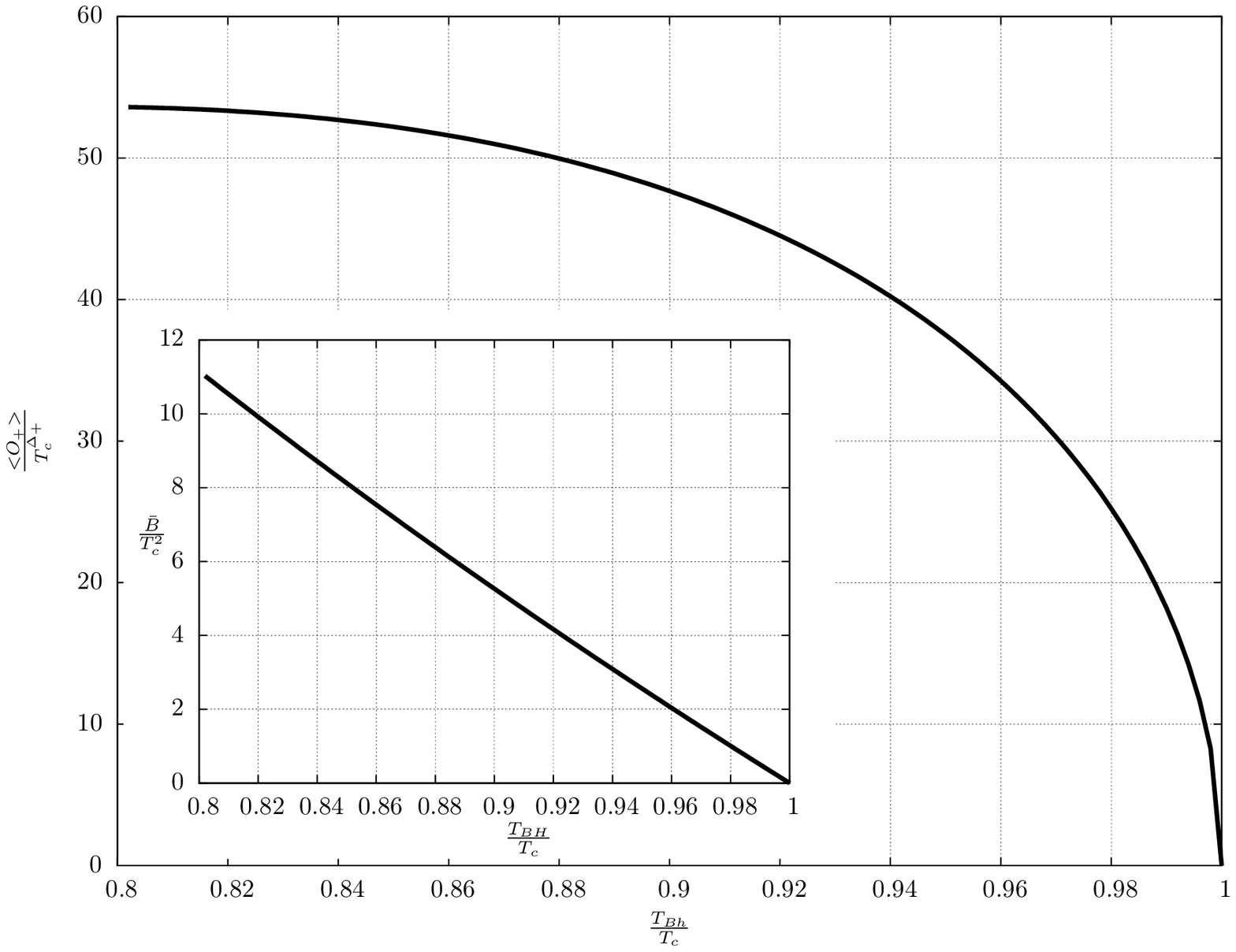}
\caption{Normalized expectation value of operator $<O_{+}>$ and critical magnetic field as the functions of the 
dimensionless temperature. Spacetime dimension is $n=5$ and matching point $z_{m} = 0.5283$. 
Remaining parameters of the scalar field are $q = 1, ~~ m^2 = -3, ~~ \Delta_{+} = 3$. Gauge field
parameters are $k = 0$ and $\alpha = 0.5$. The insert shows the behavior of the critical magnetic field
close to $T_c$.}
\label{fig2}
\end{figure}

In Fig.(\ref{fig1}) we plotted the dependence of the normalized expectation value of the condensation 
operator $<O_{+}>$ on $\frac{T_{BH}}{T_{c}}$. The plot was obtained  
for $n=4$-dimensional spacetime, for the matching point
$z_m = 0.344993$, while the parameters of scalar field were set to be $q=1,~m^2=-2,~\Delta_+=2$. The gauge field
was characterized by $k=0$ and $\alpha=0.5$. On the other hand, in Fig.(\ref{fig2}) we show the same
dependence for five-dimensional spacetime with $z_m = 0.5283$ and the scalar field parameters
$q=1,~m^2=-3,~\Delta_+=3$. The gauge field parameters are the same as in Fig.(\ref{fig1}).
One can observe that the qualitative nature of the graphs of the expectation value of the condensation operator 
remains the same for all values of $z_m$ and spacetime dimensions. Namely, the normalized 
value of the condensation operator $<O_{+}>$ behaves near $\frac{T_{BH}}{T_{c}} \rightarrow 1$ as 
$<O_{+}> \sim \sqrt{1 - \frac{T_{BH}}{T_{c}} }$. It resembles the mean field behavior of the holographic condensates
and confirms a second order kind of transitions, because the critical exponent is $1/2$.
Moreover, it turns out that the bigger spacetime dimension we consider, the greater value of the 
normalized expectation value of the operator in question we get, and the bigger value of the midpoint 
$z_m$ one takes into account.

\par
In addition in the Figs.(\ref{fig1}) and (\ref{fig2}) we show the temperature dependence of the critical magnetic field 
$\bar{B}/{T_c}^2$ on $T_{BH}/T_c$, for the same parameters as for the condensation operator $<O_{+}>$. The tendency of the 
behavior of the critical magnetic field is the same as for the normalized values $<O_{+}>$. Namely,
the critical values grow with the growth of spacetime dimensions. 
However, the qualitative nature of the dependence
is not the same, i.e., for $T_{BH} \sim T_{c}$ the 
critical magnetic field behaves as $\bar{B} \sim 1 - \frac{T_{BH}}{T_{c}}$.
The formula (\ref{expectation_value}) reveals that the coupling constant of the dark matter sector
$\alpha$ acts as a scaling factor under the square root sign. Together with the expression for
$<O_{+}>$ this indicates that the dark mater sector coupling constant should fulfill the condition that $0< \alpha < 2$, 
as $\tilde{\alpha} = 1 - \frac{\alpha^2}{4}$ ought to be positive for this expression to make sense. 

\par
As far as the other theories of gravity and other types of electrodynamics are concerned, considerations of GB-gravity
and nonlinear electrodynamics reveal that the parameter characterizing non-linearity caused the decrease of 
the critical temperature and the condensation gap greater than in Maxwell case. The increase of
GB coupling constant (characterizing the higher order curvature corrections) engenders decrease of $T_c$.
On the other hand, the behavior of the normalized value of the condensation operator $<O_{+}>$ 
indicates the second order phase transition. 
The analysis of the critical magnetic field unveils that its value grows with
the increase of non-linear electrodynamic parameter, GB-coupling constant and values of the mid-point coordinate
\cite{gre09}-\cite{kan11}.

\par
Similar behavior has been observed in our studies. The increase of dark matter coupling constant, the dimension of 
the spacetime and the value of $z_m$ causes the increase of the critical magnetic field. Consequently, the same parameters 
indicate growth of the normalized value of $<O_{+}>$.  In the light of the above studies, 
one can conclude that the effect of the dark matter sector on the holographic superconductors  plays  similar 
role as higher curvature corrections in GB-gravity and the non-linearity of the electrodynamics in question.

Finally we comment on the dependence of $\bar{B}$ on $T_{BH}/T_{c}$ as shown in the inserts to the Figs.(1) and (2). 
Close to $T_c$ the dependence is linear in agreement with the Ginzburg-Landau result \cite{ginzburg1950}.  Assuming the
validity of the standard relation between the slope \cite{degennes1964} of the critical field $(d\bar{B}(T)/dT)_{T_c}$ 
and the value of the upper critical field at zero temperature $\bar{B}(0)$
\be
\bar{B}(0)\approx 0.69~ T_c\left(\frac{d\bar{B}(T)}{dT}\right)_{T_c},
\ee  
we can evaluate zero temperature value of the upper critical field. However, the other behavior 
seen in the inserts is more interesting than that and it provides a marked 
fingerprint of the strong coupling approach. Namely, it is the upward curvature seen in the temperature dependence 
of $\bar{B}(T)$,
especially well visible in the Fig.(2), making the value of zero temperature magnetic 
field higher than the estimate based on the above equation. 
The upward curvature, which is observed in many high temperature superconductors in the standard
approach \cite{klemm1975} requires {\it inter alia} strong potential or spin-orbit scattering. 

% \cite{gurevich2003}.
 
%%%%%%%%%%%%%%%%%%%%%%%%%%%%%%%%%%%%%%%%%%%%%%%%%%%%%%%%%%%%%%%%%%%%%%%%%%%%%%%%%%%%%%%%%%%%%%%%%%%%%%%
%%%%%%%%%%%%%%%%%%%%%%%%%%%%%%%%%%%%%%%%%%%%%%%%%%%%%%%%%%%%%%%%%%%%%%%%%%%%%%%%%%%%%%%%%%%%%%%%
%\begin{appendix}

%\section{Irred   } 
%\label{irtf}
%\end{appendix}
\begin{acknowledgments}
{\L}N was supported by the Polish National Science Centre under doctoral scholarship number $2013/08/T/ST2/00122$.   
He is also grateful to Professor Piotr Chru{\'s}ciel for hospitality during his visit at the University of Vienna.   
MR was partially supported by the grant of the National Science Center $DEC-2013/09/B/ST2/03455$.
\end{acknowledgments}
%%%%%%%%%%%%%%%%%%%%%%%%%%%%%%%%%%%%%%%%%%%%%%%%%%%%%%%%%%%%%%%%%%%%%%%%%%%%%%%%%%%%%%%%%%
%%%%%%%%%%%%%%%%%%%%%%%%%%%%%%%%%%%%%%%%%%%%%%%%%%%%%%%%%%%%%%%%%%%%%%%%%%%%%%%%%%%%%%%%%%%%%%%%%%%%%%%
%%%%%%%%%%%%%%%%%%%%%%%%%%%%%%%%%%%%%%%%%%%%%%%%%%%%%%%%%%%%%%%%%%%
%%%%%%%%%%%%%%%%%%%%%%%%%%%%%%%%%%%%%%%%%%%%%%%%%%%%%%%%%%%%%%%%%%%%%%%%%%%%%%%%%
%%%%%%%%%%%%%%%%%%%%%%%%%%%%%%%%%%%%%%%%%%%%%%%%%%%%%%%%%%%%%%%%%%%%%%%%%%%%%%%%%

%%%%%%%%%%%%%%%%%%%%%%%%%%%%%%%%%%%%%%%%%%%%%%%%%%%%%%%%%%%%%%%%%%%%%%%%%%%%%%%%%%%%%%%%%%%%%%%%%%%%%%%%%%%%%%
%%%%%%%%%%%%%%%%%%%%%%%%%%%%%%%%%%%%%%%%%%%%%%%%%%%%%%%%%%%%%%%%%%%%%%%%%%%%%%%%%%%%%%%%%%%%%%%%%%%%%%%%%%%%%

%%%%%%%%%%%%%%%%%%%%%%%%%%%%%%%%%%%%%%%%%%%%%%%%%%%%%%%%%%%%%%%%%%%%%%%%%%%%%%%%%%%%%%%%%%%%%%%%%%%%%%%%%%%%%
%%%%%%%%%%%%%%%%%%%%%%%%%%%%%%%%%%%%%%%%%%%%%%%%%%%%%%%%%%%%%%%%%%%%%%%%%%%%%%%%%%%%%%%%%%%%%%%%%%
\end{document}